\documentclass[4pt, a4paper,prd,twoside,superscriptaddress, nofootinbib, twocolumn]{revtex4}
\usepackage{graphics}
\usepackage[pdftex]{graphicx}
\usepackage{amsmath, amssymb}
\usepackage{color}
\usepackage[usenames,dvipsnames]{xcolor}
\makeatletter
\newcommand*{\rom}[1]{\expandafter\@slowromancap\romannumeral #1@}
\makeatother

\usepackage{setspace}
\usepackage{setspace}
\usepackage{amsmath}
\usepackage{latexsym}
\usepackage{amssymb}
\usepackage{graphicx,color}
\usepackage{epstopdf}
\usepackage{graphicx}
\usepackage{setspace}
\usepackage{makeidx} 
\usepackage{lmodern}
\usepackage[showframe=false]{geometry}
\usepackage{enumitem}

\usepackage[T1]{fontenc}
\usepackage[utf8]{inputenc}
\usepackage{color}
\usepackage{booktabs}
\usepackage{units}
\usepackage{amsmath}
\usepackage{amssymb}
\usepackage{esint}

\pdfpageheight\paperheight
\pdfpagewidth\paperwidth

\usepackage[unicode=true,pdfusetitle, bookmarks=true,bookmarksnumbered=false,bookmarksopen=false, breaklinks=false,pdfborder={0 0 0},backref=false,colorlinks=true]{hyperref}
\hypersetup{citecolor=blue,filecolor=blue,linkcolor=blue,urlcolor=blue}

\newcommand{\qu}[1]{``#1''} 

\def\bea{\begin{eqnarray}}  
\def\eea{\end{eqnarray}}

\def\be{\begin{equation}}
\def\ee{\end{equation}}

\def\H0{H_{0}}

\def\f''{f''}

\def\q''{q''}

\def\etaf''{\eta_{f''}}

\def\etaq''{\eta_{q''}}

\usepackage{cancel}

\def\stuck{\text{St\"uckelberg}}
\makeatletter

\newcommand{\Rmnum}[1]{\expandafter\@slowromancap\romannumeral #1@}
\makeatother

\usepackage{soul}
\usepackage{anysize}
\marginsize{2.5cm}{2cm}{1cm}{3.5cm}

\usepackage{natbib}
\begin{document}


\title{A note on classical and quantum unimodular gravity}
\author{Antonio Padilla}
\email{antonio.padilla@nottingham.ac.uk}
\affiliation{School of Physics \& Astronomy, University of Nottingham, Nottingham, NG7 2RD, United Kingdom}
\author{Ippocratis D. Saltas}
\email{ippocratis.saltas@nottingham.ac.uk}
\affiliation{School of Physics \& Astronomy, University of Nottingham, Nottingham, NG7 2RD, United Kingdom}


\hoffset = -1cm
\textwidth = 19cm

\begin{abstract}
We discuss unimodular gravity at a classical level, and in terms of its extension into the UV through an appropriate path integral representation. Classically, unimodular gravity is locally a gauge fixed version of General Relativity (GR), and as such it yields identical dynamics and physical predictions. We clarify this and explain why there is no sense in which it can \qu{bring a new perspective} to the cosmological constant problem. The quantum equivalence between unimodular gravity and GR is more of a subtle question, but we present an argument that suggests one can always maintain the equivalence up to arbitrarily high momenta. As a corollary to this, we argue that whenever inequivalence is seen at the quantum level, that just means we have \emph{defined} two different quantum theories that happen to share a classical limit. We also present a number of alternative formulations for a covariant unimodular action, some of which have not appeared, to our knowledge, in the literature before.
\end{abstract}


\maketitle


\section{Introduction}
When Einstein laid down the foundations for GR \cite{GR}, he remarked that the laws of gravity sometimes took on a simpler form in certain coordinate systems, and illustrated his point by choosing so-called unimodular coordinates, where
$
\det g_{\mu\nu}=-1.
$
Of course, this choice of coordinates yields the same predictions as any other in a diffeomorphism invariant theory - a seemingly obvious fact that is at the heart of the equivalence between classical GR and so-called \emph{unimodular gravity}. 
%

Unimodular gravity is obtained from a restricted variation of the Einstein-Hilbert action, in which the condition $\det g_{\mu\nu}=-1$ is imposed from the beginning. The resulting field equations correspond to the traceless Einstein equations, and  can easily be shown to be equivalent to the full Einstein equations with a cosmological constant term, $\Lambda$, entering as an integration constant. Thus the equivalence to classical GR is made manifest, and there can be no sense in which unimodular gravity can say anything more or less than GR about anything to do with classical gravity. This includes the cosmological constant problem which is sometimes used as motivation for studying unimodular gravity \cite{vanderBij:1981ym,Buchmuller:1988wx,Buchmuller:1988yn} (see also e.g. \cite{UM1,UM2,UM3} ). One of the purposes of this paper is to make this point abundantly clear in a self-contained presentation, in the hope of addressing certain misconceptions that continue to appear in the literature.

Beyond classical gravity, however, the equivalence between GR and unimodular gravity is more subtle, with little consensus. For example, in Ref. \cite{Fiol:2008vk} it was claimed that the two theories are equivalent at the perturbative level for asymptotically flat space times, but inequivalence was found for semiclassical non--perturbative quantities around particular backgrounds.  In the canonical approach to quantum gravity, it has been suggested that unimodular gravity can help address the problem of time \cite{Unruh,Unruh-wald}, although such a claim has been strongly refuted \cite{Kuchar}. We will  present an argument suggesting that both theories can be extended into the quantum realm in such a way as  to preserve their equivalence.

Intuitively, it is straightforward to see how the equivalence can be preserved. In the path integral formalism one must always divide out the symmetry group of the theory. For GR, the symmetry group is the diffeomorphism group (Diff), 
$\delta g_{\mu\nu} =\nabla_\mu \xi_\nu+\nabla_\nu \xi_\mu$, whereas for unimodular gravity the unimodularity condition breaks this down to transverse diffeomorphisms (TDiff), satisfying $\nabla_\mu \xi^\mu=0$. One might imagine taking the path integral for GR and first dividing out the longitudinal diffeomorphsims satisfying $\nabla_\mu \xi^\mu \neq 0$, such as to give us the path integral for unimodular gravity. This is essentially the spirit behind the claims made in Ref. \cite{Fiol:2008vk}, and we are certainly sympathetic to their approach.  

Alternatively, we can always break the quantum equivalence between GR and unimodular gravity by force by \emph{defining} them to be different from the beginning. For example, one can write unimodular gravity as a manifestly Diff (rather than TDiff) invariant theory by introducing extra ($\stuck$) fields \cite{Kuchar,Henneaux}. If the extra fields are source-free the equivalence to GR remains (at least classically), but if they are sourced, it is broken. This means that the  equivalence can always be broken at the quantum level by allowing the additional fields to exist as external legs in Feynman diagrams.  



\section{Classical unimodular gravity} \label{sec:Classical-UM1}
Unimodular gravity is obtained from the Einstein-Hilbert action under a restricted variation that preserves the metric determinant, 
$
\frac{\delta}{\delta g_{\mu \nu}}  \sqrt{-g} = 0. 
$
where $g \equiv \det g_{\mu \nu} $.  To understand the implications of the unimodularity condition (UMC)  for the gauge symmetries of the theory, recall  that GR is invariant under diffeomorphism transformations, infinitesimally described by
$\delta g_{\mu \nu}(x)  =  \nabla_{\mu}\xi_{\nu} +  \nabla_{\nu}\xi_{\mu}$. 

If we think of the metric variation in the UMC as an infinitesimal, gauge transformation, one can see that the gauge vectors $\xi_\mu$ are forced to satisfy the following transversality condition $\frac{1}{2} \sqrt{-g}g_{\alpha \beta}\delta g^{\alpha \beta} = \nabla_{\mu}\xi^{\mu} = 0$, restricting the set of allowed transformations to the subset of the transverse ones (TDiff).
%


The classical equivalence between GR and unimodular gravity stems from the fact that in GR one can always choose coordinates that are unimodular, at least locally. Globally the situation is more subtle: if coordinates are fixed on the boundary, or there is no boundary, the global average $\langle \sqrt{-g}\rangle=\frac{\int \sqrt{-g} d^4 x}{\int d^4 x}$ is gauge invariant\footnote{We thank Kurt Hinterbichler for reminding us about this.}. Otherwise we can even fix the coordinate system globally and set $\langle \sqrt{-g}\rangle=1$.

The restricted variation is most conveniently imposed using a scalar Lagrange multiplier $\lambda(x)$, so that the action is given by \cite{Buchmuller:1988yn,Buchmuller:1988wx}
\be
S = \int d^{4}x \left[ \sqrt{-g} \frac{R[g]}{16 \pi G}  - \lambda(x)(\sqrt{-g} - \epsilon_0 ) \right]+S_m\label{EH+lambda}
\ee 
where $S_m$ denotes the effective action for the (quantum) matter fields coupled to the (classical) metric, and $\epsilon_0$ is a non-dynamical volume element that explicitly breaks Diff down to TDiff.  The resulting field equations yield
$G_{\mu\nu}=8 \pi G T_{\mu\nu}-\frac{\lambda(x)}{2} g_{\mu\nu}$, and $\sqrt{-g} =\epsilon_0,$
where $T_{\mu\nu}=-\frac{2}{\sqrt{-g}} \frac{\delta S_m}{\delta g^{\mu\nu}}$ is the effective energy-momentum tensor describing the matter fields. Taking the trace of {the Einstein equations} yields $\lambda(x)=\frac12 (R+8 \pi G T)$, and the traceless Einstein equations  follow, as expected. Furthermore, if we assume that the effective matter action is invariant under Diff, we have energy-momentum conservation, and  taking the divergence of the Einstein equations yield $\partial_\mu\lambda=0$. This fixes the Lagrange multiplier to be a constant $\lambda_0$, so that the dynamics is equivalent to that of GR with a cosmological constant,  $\lambda_0/2$.
Because the cosmological constant enters as an integration constant, rather than a parameter in the action, it is often said that this brings a new perspective to the cosmological constant problem. Such statements are wholly nugatory, and fail to appreciate the true nature of the problem which is one of radiative instability within effective field theory. As a result, we will  focus on an effective-field theory framework,  with quantum matter fields coupled to classical gravity, reflecting the regime in which the cosmological constant problem is most clearly posed. Note that  the details of the (unknown) full UV description of gravity are not relevant to the discussion, at least not if we wish to retain our faith in effective field theory.

The cosmological constant problem is usually described as follows: why is the observed value of Einstein's cosmological constant at least sixty orders of magnitude less than that expected from vacuum energy contributions? In quantum field theory, the vacuum is well known to carry a non-trivial energy density, and a standard calculation reveals this to be at least $\rho_{vac} \gtrsim \text{(TeV)}^4$. In the absence of gravity, one can simply define this as the zero point energy, and then ignore it as it does not enter the dynamics. When gravity is turned on,  a combination of general covariance and the equivalence principle require that this vacuum energy should, like any other form of energy, gravitate. However, unlike more familiar sources of matter such as dust or radiation, the energy density of the vacuum stays constant in time and does not dilute with the expansion of the universe. In GR,  the vacuum energy is combined with the bare cosmological constant $\Lambda$, so that it is the  combination,  $\frac{\Lambda}{8\pi G}+\rho_{vac}$, that actually gravitates.  This combination should not exceed the critical density of the universe today,  $\frac{\Lambda}{8\pi G}+\rho_{vac} \lesssim \text{(meV)}^4$, requiring  $\Lambda$ to be fine-tuned to at least sixty decimal places.

The cosmological constant problem, as described above, is somewhat wrongly stated. As mention above and discussed in some detail in \cite{seq2}, the issue is not so much one of fine-tuning, but of radiative instability. In quantum field theory, one regularly cancels off divergences in physical parameters  before fixing any finite remainder empirically using observation. Indeed from the Wilsonian Renormalisation Group (RG) we know that the UV sensitivity of relevant operators renders them incalculable -- they should be {\it measured} instead.  This is just renormalisation in action - you pick the (scale dependent) finite part of your counterterms to fit the observation. In a Wilsonian context, the latter defines the running of the corresponding operator with energy, supplemented with an appropriate renormalisation condition from observations.   The real concern is when this renormalisation procedure becomes unstable against changes in the effective description, e.g. against additional loop corrections, or under changing the renormalisation group scale in the Wilsonian effective action, a way of efficiently taking into account different quantum degrees of freedom relevant at different energies. In other words, by adding, say, additional loops, do I need to drastically retune the finite part of the appropriate counter term (in this case, the bare cosmological constant)?
In the Standard Model of Particle Physics, one typically finds that each additional loop correction adjusts the vacuum energy density by an amount $\Delta \rho_{vac} \gtrsim \text{(TeV)}^4$, requiring  the bare value of the cosmological constant to be {\it retuned} to the same  level of precision.  Similarly, in the Wilsonian action, the vacuum energy jumps by an amount $\Delta \rho_{vac} \gtrsim m^4$ whenever the cut-off passes through a threshold of mass, $m$, at least up to the TeV scale (for a nice discussion, see also \cite{cliff}). Again this requires the bare cosmological constant to be retuned to considerable precision.

Now, whether we are working with the bare parameter $\Lambda$ from GR, or the integration constant $\lambda_0$ from unimodular gravity, the essence of the cosmological constant problem remains the same. To see this,  imagine we define the effective action for matter (eg. by specifying the order in (matter) loops in a perturbative description, or else up to some  cut-off in the exact Wilsonian description) and compute the vacuum energy accordingly.  We then tune $\Lambda$ or $\lambda_0$ to some high degree of precision. But what happens when the effective description is altered slightly (e.g., by changing the loop order in the perturbative case, or the location of the cut-off beyond a new mass threshold in the non-perturbative case)? A new computation of the vacuum energy yields a completely new value, and  we are required to readjust $\Lambda$ or $\lambda_0$. In other words, a choice of $\Lambda$ in GR, or $\lambda_0$ in unimodular gravity is unstable against changing the effective field theory description of matter. Unimodular gravity does not bring any new perspective to the cosmological constant problem in comparison to GR\footnote{AP is indebted to Nemanja Kaloper for extensive discussions on this point.}.

 The simplest way to understand the cosmological constant problem is in the framework of classical gravity sourced by quantum matter fields, therefore the above discussion made at the level of the semi-classical gravity equations should be enough to convince the reader of the main argument \footnote{It is not hard to see that in the case where gravity is also treated quantum--mechanically the problem persists, and the discussion above can be generalised straightforwardly for that case.}.  However, one might still wonder how this manifests itself at the level of the action where one   might  argue that if $\sqrt{-g}$ is fixed, the cosmological constant is not a coupling of a dynamical operator in the action, and so no quantum fluctuations of any field can affect its value. However,  the point is that the restriction on $\det g$ ought to be carefully implemented, and this is most efficiently achieved via a Lagrange multiplier. Once this is done properly radiative corrections from the quantum matter Lagrangian shift this Lagrange multiplier by an overall constant, rendering its boundary value radiatively unstable. It turns out that the cosmological constant is precisely this (constant) boundary value, and so indeed the cosmological constant problem is seen to emerge just as it does in GR. See, for example, \cite{Buchmuller:1988yn}
 
 Another way to implement the constraint $|\det g|=1$ is to write  the  constrained metric, $g_{\mu\nu}$ in terms of an unconstrained metric, $f_{\mu\nu}$, where $g_{\mu\nu}=\frac{f_{\mu\nu}}{|\det f|^{1/4}}$. The renormalised cosmological constant   now enters the action  as $(\text{constant}) \times  \int d^4 x$, which one might erroneously interpret as non-dynamical. However, by a simple change of coordinates we see that this does give dynamics because, in the absence of full diffeomorphism invariance, it depends explicitly on a dynamical Jacobian\footnote{Let $x^\mu \to X^\mu(x)$, then $\int d^4 x \to \int d^4 x \left| \frac{\partial X}{\partial x} \right|$, and $\frac{\delta }{\delta X^\mu}  \int d^4 x \left| \frac{\partial X}{\partial x} \right| \neq 0$ (see  \cite{Kuchar}). }.


Let us conclude this section by presenting some alternative formulations of unimodular gravity, all of which are classically equivalent. The first of these involves restoring the  full diffeomorphism invariance  in the action (\ref{EH+lambda}) by means of a $\stuck$ trick. To this end we introduce four St\"uckelberg fields $\phi^{\alpha}(x)$, as if we were performing a  general coordinate transformation, and let $x^\alpha \to \phi^\alpha(x)$. The gravitational part of the action becomes \cite{Kuchar}, 
\be
S_\text{stuck} = \int d^{4}x \left[ \sqrt{-g} \frac{R}{16 \pi G}  -  \lambda \left( \sqrt{-g} - \epsilon_0\left| J^{\alpha}{}_{\beta}\right| \right) \right], \label{EH+lambda+phi}
\ee
where we have defined the determinant of the Jacobian matrix $J^{\alpha}{}_{\beta}\equiv  \frac{\partial \phi^{\alpha}(x)}{\partial x^{\beta}} $ as 
$
\left| J^{\alpha}{}_{\beta}\right| = 4! \delta^{[\alpha}_{\mu}\delta ^{ \beta}_\nu \delta^{ \gamma}_\kappa \delta^ {\delta]}_{\lambda} J^{\mu}{}_{\alpha} J^{\nu}{}_{\beta} J^{\kappa}{}_{\gamma} J^{\lambda}{}_{\delta}. \label{Jab-def}
$
This is now explicitly invariant under diffeomorphisms $x^{\mu} \to x'^{\mu}(x^{\nu})$, provided the $\stuck$ fields, $\phi^\alpha$, transform as scalars\footnote{Actually, the action (\ref{EH+lambda+phi}) is invariant under Diff as long as $\phi^\alpha(x) \to \Phi^\alpha(\phi(x'))$, where $\left|\frac{\partial \Phi^\alpha}{\partial \phi^\beta}\right|=1$.}. Furthermore, if we note that $\left| J^{\alpha}{}_{\beta}\right| =\partial_{\alpha} \left[ 4! \delta^{[\alpha}_{\mu}\delta ^{ \beta}_\nu \delta^{ \gamma}_\kappa \delta^ {\delta]}_{\lambda}  \phi^\mu J^{\nu}{}_{\beta} J^{\kappa}{}_{\gamma} J^{\lambda}{}_{\delta} \right]$
we see that the $\stuck$ action is a special case of the Henneaux-Teitelboim action \cite{Henneaux}, 
\be
S_\text{HT} = \int d^{4}x \left[ \sqrt{-g} \frac{R}{16 \pi G}  -  \lambda \left( \sqrt{-g} - \partial_\mu \tau^\mu\right) \right], \label{HT}
\ee
where $\tau^\mu$ is a vector density. This action can be further generalised to
\be
S_\text{genHT} = \int d^{4}x \sqrt{-g}  \left[ \frac{R}{16 \pi G} - \lambda f\left(\frac{\partial_\mu \tau^\mu}{\sqrt{-g}}\right)-q\left(\frac{\partial_\mu \tau^\mu}{\sqrt{-g}}\right) \right]. \label{genHT}
\ee
Assuming matter only couples directly to the metric, the generalised action (\ref{genHT}) gives rise to the following field equations 
\begin{align}
& G_{\mu \nu} =8\pi G\left[T_{\mu\nu}+  g_{\mu \nu} \left( \lambda V(\psi) + U(\psi) \right)\right], \label{MetricEom} \\
& f(\psi)  = 0, \;\; \;\; \partial_{\alpha}\left( \lambda f'(\psi) + q'(\psi) \right)= 0, \label{phiEom}
\end{align}
where $\psi= \frac{\partial_\mu \tau^\mu}{\sqrt{-g}}$,  $V(\psi) \equiv \psi f'(\psi) - f(\psi),~ U(\psi) \equiv \psi q'(\psi) - q(\psi)$. 
Equations (\ref{phiEom}) are constraint equations 
yielding some constant solution for $\psi=\psi_0$, and  an arbitrary Lagrange multiplier $\lambda_0= \text{constant}$.  When plugged into Einstein's equations (\ref{MetricEom}) this gives a constant, but arbitrary cosmological constant type term $ \lambda V(\psi) + U(\psi) $ on the RHS. Thus we recover the field equations of GR with a cosmological constant, as anticipated, provided that the constraint equations in (\ref{phiEom}) remain non-trivial. The latter is ensured provided the  functions $f$ and $q$ \emph{do not} fall into any of the following categories: (i) $f$ has no real zeroes (ensures that the first constraint equation in (\ref{phiEom}) has a solution for $\psi$); (ii) the isolated zeroes of $f$ and $f'$ coincide (ensures that the term $\lambda V(\psi)$ on the r.h.s of (\ref{MetricEom}) does not vanish);  (iii) $f$ is identically zero and $q$ is linear (ensures that the second cosntraint in (\ref{phiEom}) does not vanish identically). Provided above conditions are satisied, the classical dynamics for this generalised action remains equivalent to that of GR with the cosmological constant entering as an integration constant. To our knowledge this generalised form of the unimodular action has not appeared in the literature before.  
\section{Quantum  unimodular gravity} \label{sec:part1}
In this section, we will provide a (less than rigorous) argument that unimodular gravity and GR can be extended into the quantum realm in such a way as to preserve their equivalence. To this end we start by defining the generating functional,
\be \label{pathint}
Z[J]=\int Dg_{\mu\nu}D\lambda D\tau^\mu e^{iS_{HT}[g, \tau, \lambda]+iS_\text{ext}[g,  J]}
\ee
where $S_{HT}$ denotes the Henneaux-Teitelboim action (\ref{HT})  and $S_\text{ext}
$ the coupling to external sources. This should be understood as a path integral with a cut off taken to lie somewhere below the Planck scale.  We only include the leading order contributions from heavy modes to the low energy effective action, which is assumed to be invariant under Diffs, as is the low energy functional measure\footnote{Our argument is not sensitive to the details of how we define $Dg_{\mu\nu}$ so we will not dwell on any of  subtleties  associated with the path integral approach to quantum GR. }.  Crucially,  we have assumed that it is only the metric that couples to external sources and not the vector density, $\tau^\mu$ or the scalar, $\lambda$\footnote{What we really mean here is that $\tau^\mu$ and $\lambda$ should  really be thought of as auxiliary fields, and do not correspond to asymptotic states. It does not, therefore, make sense to speak of $n$ points functions of these fields since they can only ever appear as internal lines in Feynman diagrams. This is consistent with the statement that the matter Lagrangian is Diff invariant, although it is a somewhat stronger statement. If the matter Lagrangian were not Diff invariant one could argue that matter fields sourced the Stuckleberg fields, which could in principle  be identified with $\tau^\mu$.}. Furthermore, the Henneaux-Teitelboim action has been endowed with a boundary term \cite{Fiol:2008vk}, 
$\int_{\partial V} d^3 x \sqrt{-\gamma} \left[\frac{1}{8\pi G} K-n_\mu \lambda \tau^\mu \right]$
where $\gamma \equiv \det \gamma_{\mu \nu}$ with $\gamma_{\mu\nu}$ the induced metric on the boundary, $n_\mu$ is the outward normal, and $K \equiv K^{\mu}{}_{\mu}$ is the trace of the extrinsic curvature. After integration by parts it is easy to see that $\tau^\mu$ reduces to a Lagrange multiplier whose purpose is merely to fix $\delta_\mu \lambda=0$. For a suitably chosen measure, the functional integration over $\tau^\mu$ should  yield
\be \label{Zwithdelta}
Z[J]=\int D g_{\mu\nu}D \lambda  \delta\left[ \delta_\mu \lambda \right]e^{i \bar S_{HT}[[g, \lambda]+iS_\text{ext}[g, J]}
\ee
where 
\begin{align}
 \bar S_{HT} [g, \lambda]= \int_V d^{4}x \sqrt{-g} \left[\frac{R}{16 \pi G}  -  \lambda(x) \right] 
 +\int_{\partial V} d^3 x \sqrt{-\gamma} \frac{1}{8\pi G} K
\end{align}
In \cite{Fiol:2008vk}, it is argued that a physical boundary condition would be to impose no variation of $\lambda$ at the boundary\footnote{This follows from the observation that $\lambda$ is an observable, while $\tau^\mu$ not. We refer the reader to \cite{Fiol:2008vk} for more details around this point.}. This, along with the delta function, allows us to completely   do the functional integration over $\lambda$, yielding
\be
Z[J]=\int D g_{\mu\nu}e^{i \bar S_{GR}[g; \lambda_0)+iS_\text{ext}[g, J]}
\ee
where 
\begin{multline}
\bar S_{GR} [g; \lambda_0)= \int_{V} d^{4}x \sqrt{-g} \left[\frac{R}{16 \pi G}  -  \lambda_0 \right] +\int_{\partial V} d^3 x \sqrt{-\gamma} \frac{1}{8\pi G} K
\end{multline}
and $\lambda_0$ is the arbitrary fixed boundary value of $\lambda$. Thus we arrive at the generating functional for GR with a cosmological constant $\lambda_0/2$.  Again, this path integral should be understood as being cut-off, keeping only the leading order contributions to the effective action from integrating out heavy modes above the cut-off. Our somewhat schematic argument strongly suggests that there is a clear way in which we can extend unimodular gravity in to the UV so that it maintains its equivalence to GR. 


What if we do not fix $\lambda$ on the boundary? Then the delta function in Eq. (\ref{Zwithdelta}) does not allow us to completely do the functional integration over $\lambda$. In particular we are left with an ordinary integration over space-time constants $\lambda_0$, 
\be
Z[J]=\int D g_{\mu\nu} \int d\lambda_0 f(\lambda_0) e^{i \bar S_{GR}[g; \lambda_0)+iS_\text{ext}[g, J]}
\ee
where we have included a possible non-trivial contribution, $f(\lambda_0)$,  to the measure for completeness. Classically this suggests a theory which is \emph{locally} equivalent to GR but with an additional global constraint coming from variation over a \emph{global} parameter - the bare cosmological constant (in this case, $\lambda_0/2$).  
The two subtlety different possibilities we have just described are obviously related to whether or not one is able to fix the unimodular gauge globally in GR, as discussed in the previous section.
%

We expect to break above equivalence the moment we switch on corresponding sources, or in other words, we allow $\lambda(x)$ and $\tau^\mu(x)$ to lie on the external legs of Feynman diagrams. However, we emphasise that by doing this we are breaking the quantum equivalence {\it by hand}.
\section{Discussion and conclusions} \label{sec:Conclusions} 
There are three messages we would like the reader to take away from this paper:
\begin{enumerate}[leftmargin=*]
\item Classical unimodular gravity $=$ classical GR, so any suggestion that the former can shed new light on any problems faced by the latter are entirely nugatory.
\item Quantum unimodular gravity $=$ quantum GR provided we make certain assumptions about how we extend into the UV.
\item Quantum unimodular gravity $\neq$ quantum GR if we break those assumptions, but that is our choice, and amounts to \emph{defining} the theories to differ in the UV. 
\end{enumerate}

The classical equivalence and its implications are spectacularly obvious, but confusion continues to reign in the literature. The quantum equivalence between the two theories is more of a subtle issue.  We have presented a schematic argument based on the path integral approach to quantum gravity that suggests one can always maintain equivalence up to arbitrarily high momenta. The argument uses covariant  descriptions of unimodular gravity \cite{Henneaux,Kuchar}, where additional fields can be rendered purely auxiliary such that they may be integrated out in the path integral leaving us with the path integral for GR, with appropriate boundary conditions.

Whenever the quantum equivalence is seen to fail, we would argue that this says more about how one chose to go about extending the theories into the UV, than some inevitable inequivalence at the quantum level. Indeed, that choice amounts to \emph{defining} the quantum theories to be inequivalent.  

To sum up then: classical unimodular gravity and classical GR are the same thing, and they can be extended into the UV such that the equivalence is maintained. Whenever inequivalence is seen at the quantum level, that just we means we have \emph{defined} two different quantum theories that happen to share a classical limit. An example of the latter is given by \cite{Kuchar, Saltas:2014cta} with the $\stuck$ fields and Lagrange multiplier \emph{allowed to lie on external legs}.

\acknowledgements
We would like to thank Nemanja Kaloper, Kiril Krasnov, Kurt Hinterbichler, Paul Saffin, Iggy Sawicki and David Stefanyszyn for interesting discussions. AP is supported by a Royal Society URF, and IDS by an STFC consolidating grant.

\bibliography{UM}
\end{document}